\documentclass{article}
\usepackage{ijcai23}

\usepackage{times}
\usepackage{soul}
\usepackage{url}
\usepackage[hidelinks]{hyperref}
\usepackage[utf8]{inputenc}
\usepackage[small]{caption}
\usepackage{graphicx}
\usepackage{amsmath}
\usepackage{amsthm}
\usepackage{booktabs}
\usepackage{algorithm}
\usepackage[switch]{lineno}

\usepackage{csquotes}
\usepackage{amsfonts}
\usepackage{bm}
\usepackage{pythonhighlight}
\lstnewenvironment{pythonLines}[1][]{\lstset{style=mypython,numbers=left}}{}

\title{Efficient Data Mosaicing with Simulation-based Inference}

\author{
Andrew Gambardella$^1$
\and
Youngjun Choi$^1$\and
Doyo Choi$^1$\And
Jinjoon Lee$^1$
\affiliations
$^1$Graduate School of Culture Technology, KAIST
\emails
\{atgambardella, youngjun.choi, doyochoi, jinjoon.lee\}@kaist.ac.kr
}

\begin{document}

\maketitle

\begin{abstract}
    We introduce an efficient algorithm for general data mosaicing, based on the simulation-based inference paradigm. Our algorithm takes as input a target datum, source data, and partitions of the target and source data into fragments, learning distributions over averages of fragments of the source data such that samples from those distributions approximate fragments of the target datum. We utilize a model that can be trivially parallelized in conjunction with the latest advances in efficient simulation-based inference in order to find approximate posteriors fast enough for use in practical applications. We demonstrate our technique is effective in both audio and image mosaicing problems.
\end{abstract}

\section{Introduction}

Among post-structuralist texts, the 1980 book ``A Thousand Plateaus: Capitalism and Schizophrenia'' by Deleuze and Guattari stands out as a seminal experimental work of philosophy, dealing with a wide range of topics originating from the natural world. On the topic of language, the authors state the following:

\begin{displayquote}
...relatively few linguists have analyzed the necessarily social character of enunciation...The social character of enunciation is intrinsically founded only if one succeeds in demonstrating how enunciation in itself implies \textit{collective assemblages}. It then becomes clear that the statement is individuated, and enunciation subjectified, only to the extent that an impersonal collective assemblage requires it and determines it to be so...every statement of a collective assemblage of enunciation belongs to indirect discourse...Direct discourse is a detached fragment of a mass and is born of the dismemberment of the collective assemblage; but the collective assemblage is always like the murmur from which I take my proper name, the constellation of voices, concordant or not, from which I draw my voice~\cite{Mille1980}.
\end{displayquote}

Whereas philosophers such as Deleuze describe the world in natural language, artists evoke the world itself through media. The most natural tool with which a contemporary media artist could evoke the ideas espoused in this quote would be a mosaic, directly constructing a collective assemblage of ``voices'' which materialize into a so-called novel ``subjectified enunciation'' directly from the aggregate of detached fragments of ``direct discourse.'' In more common parlance, this means finding, from a set of source data, fragments of the data which could be rearranged and overlapped so as to approximate an entirely different target datum. Such a mosaic would serve as a metaphor for how linguistic meaning is created in a social manner, and that all enunciations of natural language that can be understood must be inherited from other enunciations which others had heard and understood previously, an idea which has resonated throughout Western philosophy for centuries, notably having been used to humorous effect by Humpty Dumpty in Lewis Carroll's 1871 novel ``Through the Looking-Glass''~\cite{Carroll1871} and expanded upon in Wittgenstein's ``Philosophical Investigations''~\cite{WittgensteinInvestigations} and Davidson's ``A Nice Derangement of Epitaphs''~\cite{DavidsonEpitaphs}.

The algorithmic art community has developed a number of tools and approaches to mosaicing in many different modalities. As a further contribution to this field, in this paper we propose a generalized and data-agnostic approach to mosaicing by approaching data mosaicing as a Bayesian inference problem. Recent advances in probabilistic programming~\cite{van2018introduction} and simulation-based inference~\cite{cranmer2020frontier} allow for statisticians to write a stochastic model for mosaicing naturally in a programming language, using inference techniques to condition on a target datum and discover a posterior over traces (i.e., runs of the model) such that in the aggregate, samples from the posterior produce output which closely approximates the target datum. This process requires only the model specification and the verification of inference results, a simple task relative to the prohibitively daunting requirements imposed when creating a mosaic out of many overlapping components in an interactive computer-aided setup as is done commonly.

The main contributions of this work are as follows:
\begin{itemize}
    \item We introduce a stochastic model for data generation through mosaicing via simple averaging.
    \item We show that this model can be implemented in a probabilistic programming language and effectively conditioned on a real datum which did not originate from the model, allowing us to discover an interpretable and disentangled representation of arbitrary data as mixtures of other data in the form of a posterior distribution, from which we can sample a potentially limitless collection of mosaics for any given target datum.
    \item We show how recent advances in simulation-based inference allow for one to perform inference efficiently, creating mosaics many thousands of times faster than a naive baseline approach.
    \item We demonstrate that one singular mosaicing model can be applied to multiple modalities in a data-agnostic manner, showing numerous experiments in both audio and image mosaicing.
    \item We demonstrate that our model produces qualitatively good results even in extremely low compute regimes.
\end{itemize}


\begin{figure}
    \centering
    \includegraphics[width=0.39\textwidth]{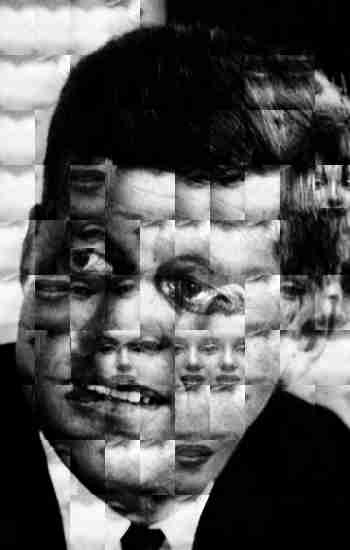}
    \caption{\textit{JFK-MM} by Adam Finkelstein and Sandy Farrier. A photographic mosaic of John F. Kennedy made from parts of Marilyn Monroe pictures which was exhibited in the Xerox PARC Algorithmic Art Show in 1994. Our method uses simulation-based inference in order to create similar mosaics using arbitrary data.}
    \label{fig:jfk}
\end{figure}

\section{Preliminaries}

\subsection{Simulation-based Inference}

\textit{Simulation-based inference} refers to a collection of techniques used to perform Bayesian inference over stochastic latent variables in a computer program. Typically these programs are written using a \textit{probabilistic programming language} which allows for this inference procedure to be done automatically, using approximate inference algorithms which are specially tuned to work over runs of computer programs. Each run of a computer program, used to form empirical distributions over its random variables, is referred to as a \textit{trace} in the probabilistic programming literature.

Simulation-based inference techniques are rooted in the theory of Approximate Bayesian Computation (ABC)~\cite{marjoram2003markov,wilkinson2013approximate}, in which the likelihood is typically not computed directly. In ABC, multiple traces of the simulator are aggregated into an empirical posterior, with traces being accepted or given high weight when their observed variables align closely with a target observation (where ``closeness'' is defined via a tolerance hyperparameter), and rejected or given low weight otherwise. Several different approximate inference algorithms can be used under this framework, but the present work examines the use of Markov chain Monte Carlo (MCMC)~\cite{wingate2011lightweight,van2018introduction} due to its verifiable convergence guarantees.

\subsection{Markov Chain Monte Carlo (MCMC)}

Markov chain Monte Carlo (MCMC) is a class of algorithms that allow for sampling from a target probability distribution which cannot be determined analytically. These algorithms all involve the construction of a Markov chain which has the target probability distribution as its equilibrium distribution, but each algorithm differs in its construction of this chain. The most basic of these algorithms, which is commonly implemented in probabilistic programming libraries, is known as \textit{random walk Metropolis-Hastings}~\cite{Metropolis1953,Hastings1970,wingate2011lightweight}, which involves finding new model parameters at each step via a random walk, and accepting or rejecting these new parameters based on how closely the model outputs match the observation relative to the previous model parameters in the chain. A much more effective, but much harder to implement, variation of Metropolis-Hastings can be found in \textit{Hamiltonian Monte Carlo} (HMC)~\cite{DuaneHMC,RadfordHMC}, which proposes new moves in state space with an approximate Hamiltonian dynamics simulator. The advantage of this approach over random walk Metropolis-Hastings is that successive moves in state space using HMC are much less correlated with previous states than they would be using random walk Metropolis-Hastings. This in turn leads to a drastic reduction in the number of forward runs of the model that are needed both during the warmup stage (i.e., before the chain has converged) and when collecting posterior samples, ultimately leading to convergence many orders of magnitude faster in terms of wall-clock time.

\section{Prior Work}

Data mosaicing, particularly photographic mosaicing, has been one of the mainstays of algorithmic art for decades. The first algorithmic photographic mosaic exhibited as art was most likely \textit{JFK-MM} by Adam Finkelstein and Sandy Farrier, reproduced in Figure~\ref{fig:jfk}. These early methods, including extensions to the video domain, relied heavily on color correction of the selected images in order to produce mosaics which appeared to be similar to the target, as well as a potentially large amount of manual human intervention in the image selection process~\cite{finkelstein1998image,KleinVideoMosaics}. In contrast, our work produces mosaics in a fully automated manner, and does not require the modification of source data in any way.

In the audio mosaicing domain, many prior tools required the specification of heavily hand-engineered features or involved intensive human labor in creating the mosaic~\cite{MusicalMosaicing,Mosievius,ColemanMosaic}. In the Bayesian audio mosaicing domain, previous algorithms utilized heavily hand-engineered audio-specific models which scaled poorly~\cite{HoffmanSpectral}. Finally, more recent methods based on deep learning also modify the source data in unpredictable ways, rendering the resulting ``mosaic'' entangled and uninterpretable~\cite{LetItBee2015}. In contrast, our method is nearly fully automated, works on data in nearly any form, is amenable to parallelization and thus scales better than previous Bayesian mosaicing methods, and is fully interpretable (in that one may inspect samples and determine exactly how they were created through mosaicing of the source data).

\section{Algorithm}
\label{sec:alg}

\begin{algorithm*}
    \caption{Numpyro (0.10.1) code for our probabilistic generative model of a target datum fragment, which is created by selecting and averaging clips from a pre-defined source collection of data fragments. This code is data-agnostic, and was used for all experiments in our paper.}

    \begin{pythonLines}
import jax.numpy as jnp
import numpyro
import numpyro.distributions as dist

def model(target, sources, timestep, stddev, num_clips):
    """Re-creates target fragment by averaging num_clips random source fragments.

    Keyword arguments:
    target -- one fragment of the target datum
    sources -- a tensor of source data broken into fragments; axis 0 indexes those fragments
    timestep -- index of the current target datum fragment
    stddev -- tolerance hyperparameter for inference
    num_clips -- hyperparameter for number of source data fragments to be mixed together
    """
    selected_source_fragments = numpyro.sample('selected_source_fragments_{}'.format(timestep), dist.Categorical(jnp.ones((num_clips, sources.shape[0]))))
    clips = sources[selected_source_fragments]
    averaged_clips = jnp.average(clips, axis=0, keepdims=False)
    numpyro.sample('obs_{}'.format(timestep), dist.Normal(averaged_clips.reshape(-1), stddev), obs=target.reshape(-1))
    \end{pythonLines}
    \label{alg:model}
\end{algorithm*}

In order to enable efficient inference, in all experiments we split our data (both source and target) into natural ``data fragments'' and perform inference on each of these fragments independently. For example, for images we split our data spatially, and for audio we split our data into Short-time Fourier Transform (STFT) segments~\cite{SignalProcessingAdvances}. These splits allow us to construct an embarrassingly parallel inference problem, performing independent MCMC runs for each fragment simultaneously on separate CPUs. We also note that the time required for probabilistic programming inference scales supralinearly with observation size, so that splitting the data into fragments and processing each fragment independently results in speed gains many times larger than the number of CPUs used in parallel.

The data-agnostic probabilistic model we used for all experiments is specified in Algorithm~\ref{alg:model}. The inputs to our model are specified in the function definition and docstring in lines 5 to 14. In line 15 we draw \textit{num\_clips} samples from a categorical distribution over the indices of the \textit{sources} tensor; the parameters of this categorical distribution are initialized to be uniform, but will be learned through inference. In line 16 we obtain those \textit{num\_clips} fragments through integer array indexing applied to the \textit{sources} tensor and in line 17 we average those fragments. In line 18 we construct an isotropic multivariate normal distribution with a mean centered at \textit{averaged\_clips}, and with standard deviation given by the hyperparameter \textit{stddev}, which can be interpreted as a tolerance hyperparameter. During MCMC inference, we will be more likely to accept traces which lead to such normal distributions under which the observed \textit{target} tensor has high probability. It is through this process that we learn parameters for the categorical distribution in line 11 that will result in mosaics that match closely with the target datum fragment.

Our algorithm is simple, natural, and easy to implement; other than the code in Algorithm~\ref{alg:model}, we only required boilerplate code for data processing and running inference. We initially implemented our algorithm in PyProb~\cite{Baydin2018,Baydin2019}, which includes algorithms for random walk Metropolis-Hastings inference in models with discrete latent variables, as we have here. We found, however, that random walk Metropolis-Hastings alone was not enough for our algorithm to be used in practical cases, as inference required about 100,000 warmup samples, even for target audio clips as short as five seconds long. We therefore re-implemented our model in NumPyro~\cite{phan2019composable,bingham2019pyro}, which has support for Hamiltonian Monte Carlo inference with the No U-Turn Sampler (NUTS)~\cite{HoffmanNUTS} in models with discrete latent variables through the \textit{DiscreteHMCGibbs} function. These more powerful inference algorithms allowed us to converge to a posterior with orders of magnitude fewer samples compared to random-walk Metropolis-Hastings. With other speedups obtained through the parallelization of our model and through JAX's JIT compilation~\cite{jax2018github}, we were able to perform significantly more complex Bayesian mosaicing experiments than had been reported in previous literature, which used models which fundamentally could not be parallelized as ours can~\cite{HoffmanSpectral}.

\section{Experiments}

In this section, we describe experiments over multiple modalities, showing the power and versatility of the simulation-based inference paradigm, and our specific model, in algorithmic art applications. We found over all experiments that our model is not sensitive to hyperparameter choices, and that differing choices of hyperparameters allowed for mosaics with different properties, but all of which qualitatively appeared to be approximating the target datum when possible. Specifically, over each of the following problems and given any combination of hyperparameters, we never witnessed our model failing to converge to a posterior (where failure would mean that multiple MCMC chains with different random seeds would arrive at differing stationary distributions~\cite{Gelman1992,Brooks1998}).

In all following experiments over all modalities, we set \textit{num\_clips}, the number of source data fragments to be averaged, to be 30. For all image mosaicing experiments we used 1000 warmup steps, and for audio mosaicing experiments we used 20000 warmup steps. All experiments were performed on an AMD Ryzen 9 3900X 12-Core processor, which provides up to 24 threads through simultaneous multi-threading.

\subsection{Image Mosaicing}

\begin{figure}
    \centering
    \includegraphics[width=0.29\textwidth]{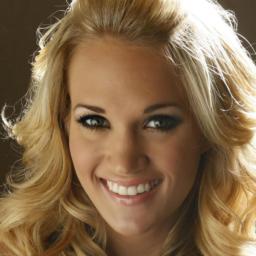}

    \caption{The final image in the CelebA-HQ dataset, which we used as the target for all of our image mosaicing experiments.}
    \label{fig:targetimage}
\end{figure}

All of our image mosaicing experiments were based on the CelebA-HQ dataset~\cite{ProgressiveGANs}, with each image preprocessed by resizing to 256x256 and normalizing the RGB values to lie in $[0, 1]$. We demonstrate two types of mosaicing with our method, ``in-place mosaicing'' and ``photographic mosaicing''. For all of our experiments in this section, we use the first 1000 images in the CelebA-HQ dataset as our source image dataset, and the final image (image \#30,000) in the CelebA-HQ dataset as our target. The target image is reproduced in Figure~\ref{fig:targetimage} for reference.

\subsubsection{CelebA-HQ In-place Mosaicing}
\label{sec:inplace}

\begin{figure}
    \centering
    \includegraphics[width=0.09\textwidth]{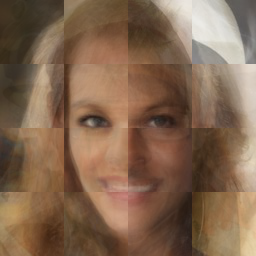}
    \includegraphics[width=0.09\textwidth]{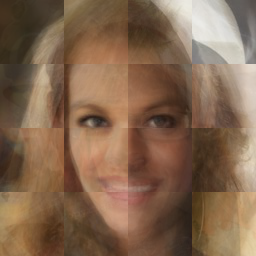}
    \includegraphics[width=0.09\textwidth]{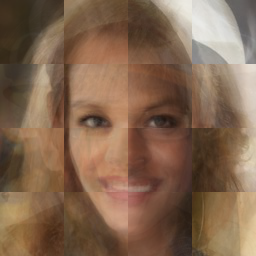}
    \includegraphics[width=0.09\textwidth]{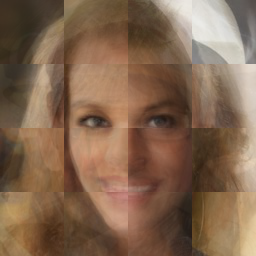}
    \includegraphics[width=0.09\textwidth]{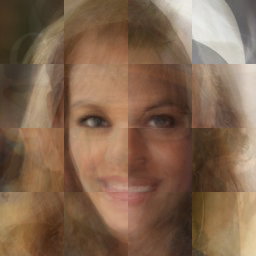}
    \includegraphics[width=0.09\textwidth]{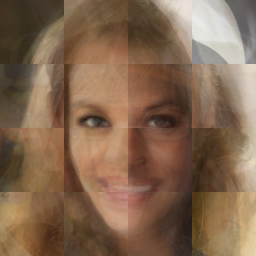}
    \includegraphics[width=0.09\textwidth]{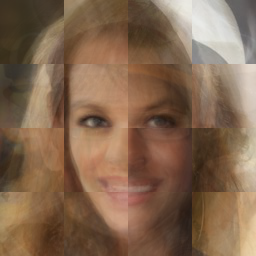}
    \includegraphics[width=0.09\textwidth]{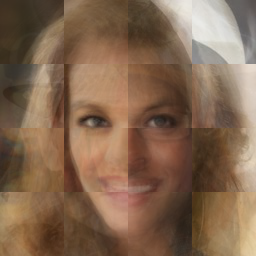}
    \includegraphics[width=0.09\textwidth]{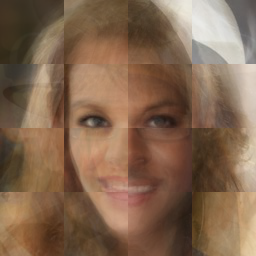}
    \includegraphics[width=0.09\textwidth]{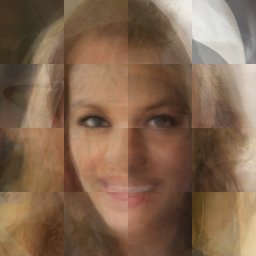}

    \caption{Re-construction of the last image in the CelebA-HQ dataset using an in-place mosaic over the first 1000 images, with a small tolerance standard deviation. This allows for some variation in samples that can be noticed when viewed closely, but each mosaic matches the target image more or less as closely as possible.}
    \label{fig:celeba_smallstddev}
\end{figure}

We first demonstrate our technique on what we refer to as ``in-place mosaicing'', where the target image is split into fragments spatially, and then each fragment is individually approximated by an assemblage of spatially corresponding fragments from the source data set. In our experiments, we split each image into 16 squares of size 64x64, and approximate each square with an average combination of 30 corresponding squares from the source dataset. In Figure~\ref{fig:celeba_smallstddev} we show samples from our model when re-constructing the last image in the CelebA-HQ dataset using the first 1000 images, with a small tolerance standard deviation \textit{stddev}. From a distance each of the samples looks very similar, but upon close inspection, one can notice minute differences. When a small standard deviation is applied, our method essentially acts like an ``averaged nearest neighbors,'' searching for \textit{num\_clips} source fragments that, when averaged, most closely approximate the target datum. While this is a useful application of our tool in and of itself, the real strength of the simulation-based inference paradigm lies in its ability to learn rich posterior distributions from which one could sample a diverse set of mosaics for any given target datum.

\begin{figure}
    \centering
    \includegraphics[width=0.09\textwidth]{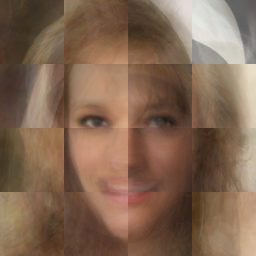}
    \includegraphics[width=0.09\textwidth]{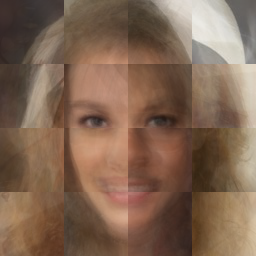}
    \includegraphics[width=0.09\textwidth]{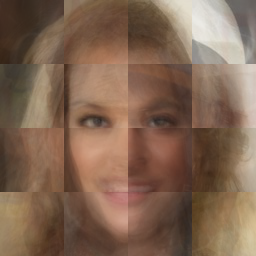}
    \includegraphics[width=0.09\textwidth]{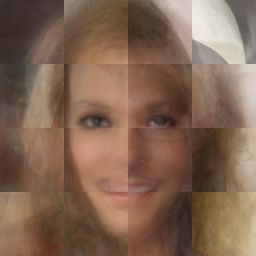}
    \includegraphics[width=0.09\textwidth]{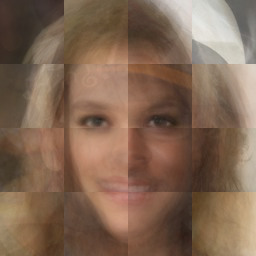}
    \includegraphics[width=0.09\textwidth]{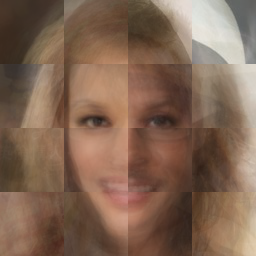}
    \includegraphics[width=0.09\textwidth]{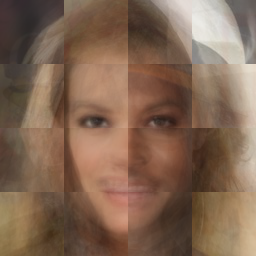}
    \includegraphics[width=0.09\textwidth]{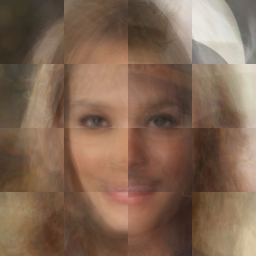}
    \includegraphics[width=0.09\textwidth]{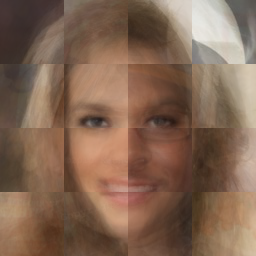}
    \includegraphics[width=0.09\textwidth]{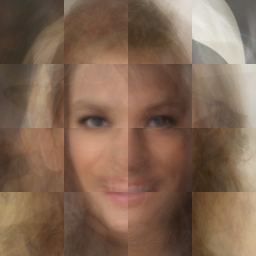}

    \caption{Re-construction of the last image in the CelebA-HQ dataset using an in-place mosaic over the first 1000 images, with a larger tolerance standard deviation. Here there is much more noticeable variation in the samples, although the general patterns still largely match the target image.}
    \label{fig:celeba_largestddev}
\end{figure}

We therefore compare with samples taken from a model with a larger tolerance standard deviation, shown in Figure~\ref{fig:celeba_largestddev}. As expected, the samples obtained here are more diverse than those with a small tolerance, but with this diversity, the samples diverge from the target image. This tradeoff between sample diversity and sample fidelity is a choice that the simulation-based inference paradigm offers to artists, and should be seen as a strong advantage of this method of mosaicing over others.

We also note that there is a clear mosaic-like effect in our output reconstructions. This is due to our choice of setting the mosaic window stride to be equal to the mosaic window size, and thus causing no overlap between the target data fragments. Alternatively, one could choose a stride and window size which allows for overlap, and reconstructions with this setup should not have this mosaic-like effect. Such choices should be determined by the end-users of these tools in order to create desired effects.

Our in-place image mosaicing experiments using 16 fragments of size 64x64 took approximately 5 hours of wall clock time to run.

\subsubsection{CelebA-HQ Photographic Mosaicing}
\label{sec:photographic}

\begin{figure}
    \centering
    \includegraphics[width=0.09\textwidth]{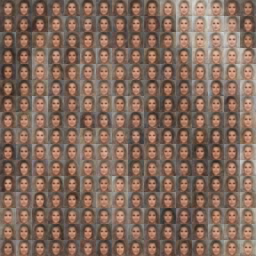}
    \includegraphics[width=0.09\textwidth]{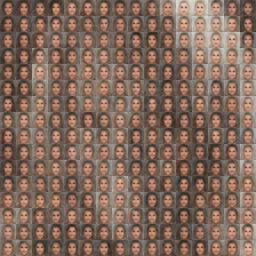}    \includegraphics[width=0.09\textwidth]{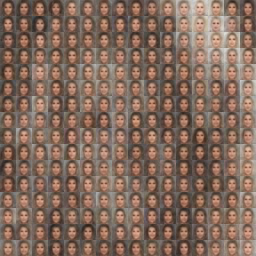}    \includegraphics[width=0.09\textwidth]{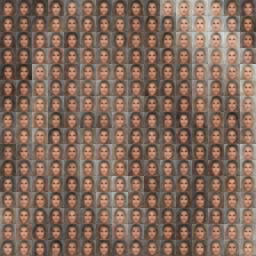}    \includegraphics[width=0.09\textwidth]{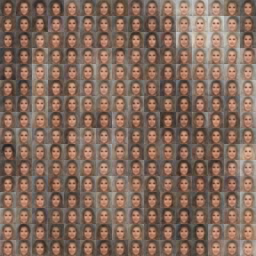}
    \includegraphics[width=0.09\textwidth]{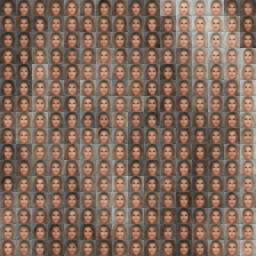}    \includegraphics[width=0.09\textwidth]{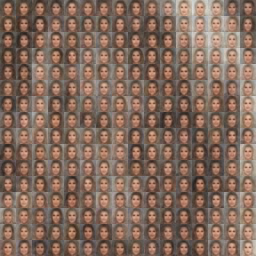}    \includegraphics[width=0.09\textwidth]{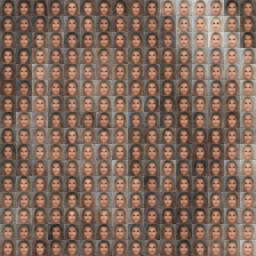}    \includegraphics[width=0.09\textwidth]{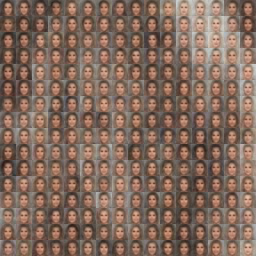}    \includegraphics[width=0.09\textwidth]{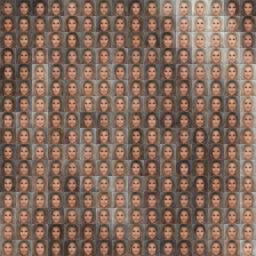}
    \caption{A photographic mosaic of the last image in the CelebA-HQ dataset using the first 1000 images downscaled to 16x16. Best viewed digitally at multiple scales of resolution.}
    \label{fig:celeba_photographic}
\end{figure}

We next demonstrate our technique on photographic mosaicing. Again we use a similar setup, attempting to approximate the final image in the CelebA-HQ dataset using the first 1000 images, but now we resize the first 1000 images down to 16x16, and allow each of those resized images to be used in any of the squares of the approximation. We show samples from our model using this setup in Figure~\ref{fig:celeba_photographic}. While from afar our photographic mosaic appears to be an approximation of the final image in CelebA-HQ (see Figure~\ref{fig:targetimage}), zooming in closely, one can see that the image is made up of 256 CelebA-like faces. Further inspecting, we can see that in the blonde hair of the approximation, the model has not only selected source images with light backgrounds, but also the faces in those images have blonde hair and paler complexions than those darker regions of the target image, such as the eyes.

We note that we attempted a similar photographic mosaicing experiment using 16 blocks of size 64x64, and while MCMC was able to converge, the resultant mosaic could not be interpreted as the target image. This seems to be due to the lack of variation in CelebA images, which always have a human face centered in the image. Failure cases such as these stress the importance of both qualitative and quantitative evaluation of simulation-based inference methods, particularly when applied to creative tasks; it is not enough for MCMC to converge, but a human must also evaluate whether the output meets their needs or not.

Our photographic image mosaicing experiments using 256 fragments of size 16x16 took approximately 1 hour of wall clock time to run. In a hypothetical distributed setting using 256 threads on CPUs of similar power, this experiment would have taken approximately 6 minutes. We note that despite the outputs in this experiment and the outputs in the in-place image mosaicing experiments being the same size (256x256 images) this experiment was much faster precisely because of the relatively small observation size (many 16x16 image fragments instead of few 64x64 image fragments); as noted earlier in the paper, time required for probabilistic programming inference scales supralinearly with respect to observation size.

\subsection{Audio Mosaicing}
\label{sec:audiomosaicing}

\begin{figure}
    \centering
    \includegraphics[width=0.48\textwidth]{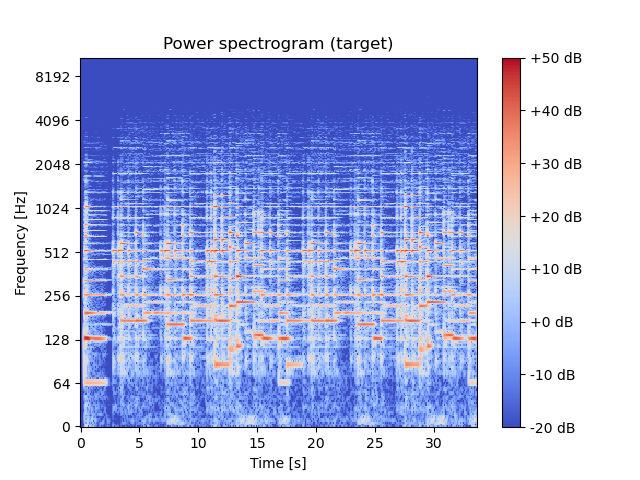}
    \includegraphics[width=0.48\textwidth]{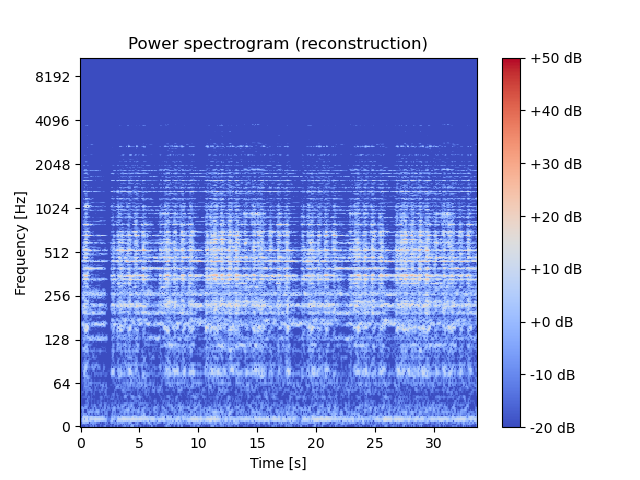}

    \caption{Spectrogram of a MIDI rendition of \textit{Happy Birthday to You} featuring multiple different instruments (top) compared with the spectrogram of a reconstruction made by mosaicing clips from a live piano rendition of \textit{Für Elise} (bottom). Samples from our model can be heard here: \url{https://soundcloud.com/datamosaicing/sets/fur-elise-happy-birthday}. The source audio can be heard here: \url{https://youtu.be/s71I_EWJk7I} and the target audio can be heard here: \url{https://youtu.be/Dug9ej1i5OY}.}
    \label{fig:happybirthday}
\end{figure}

Using the same model, we also performed audio mosaicing experiments. We preprocessed and fragmented our audio by converting it into its frequency representation through the Short-time Fourier Transform (STFT) with an FFT window size of 8192, which corresponds to 372 milliseconds of audio with a sample rate of 22050 Hz. We then further preprocessed our audio data by rescaling to decibels relative to a fixed reference, and then taking its absolute value, converting its complex frequency representation into scalar magnitudes and thereby halving the size of observed data. While this preprocessing step causes a loss of information, qualitatively we find that the task of ``matching real-valued magnitudes'' rather than ``matching complex-valued frequency representations'' still produces high quality audio mosaics. Furthermore, despite the loss of information in the preprocessing step, we are still able to perfectly reconstruct the mosaiced audio signal due to the disentangled nature of probabilistic programming posteriors allowing us to inspect which fragments were selected by the algorithm, and use their complex frequency representations during signal reconstruction via the Inverse Short-time Fourier Transform. Finally, for each target fragment, we only consider mosaicing the closest $k$ source fragments in terms of cosine distance, with $k$ a hyperparameter. We justify this by nothing that in practice, source fragments which are extremely distant from the target fragment are nearly never selected by our algorithm even if they were available for consideration, causing this preprocessing step to accelerate the convergence of our algorithm with little or no downside. For the following experiments we set $k$ to 200 (compared to the 2468 fragments in the original source data set), but any reasonable value should be acceptable, with smaller values speeding up inference time, and larger values increasing sample diversity and fidelity.

We show the spectrogram for a MIDI rendition of \textit{Happy Birthday to You}, as well as the spectrogram of a mosaic of it formed from fragments taken from a live piano rendition of \textit{Für Elise}, in Figure~\ref{fig:happybirthday}. We note that the mosaic reconstruction appears to be a ``weaker'' version of the original due to the uncertainty inherent in Bayesian posteriors, as well as the differing timbre of the source and target audio. Qualitatively, the reconstruction sounds like a rendition of \textit{Happy Birthday to You} on a grand piano.

This audio mosaicing experiment took approximately 22 hours of wall clock time to run. In a hypothetical distributed setting using 363 threads on CPUs of similar power, this experiment would have taken approximately 1.5 hours to run.

\subsection{Extremely Low Compute Mosaics}

\begin{figure}
    \centering
    \includegraphics[width=0.09\textwidth]{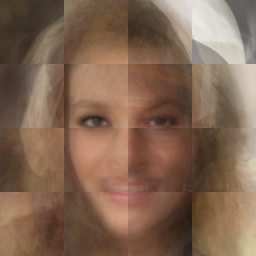}
    \includegraphics[width=0.09\textwidth]{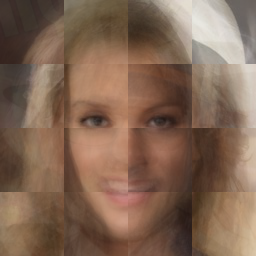}    \includegraphics[width=0.09\textwidth]{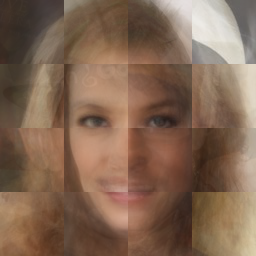}    \includegraphics[width=0.09\textwidth]{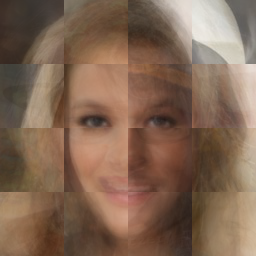}    \includegraphics[width=0.09\textwidth]{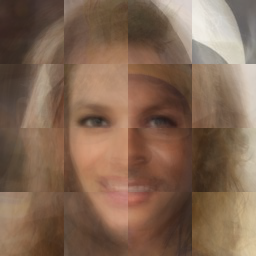}
    \includegraphics[width=0.09\textwidth]{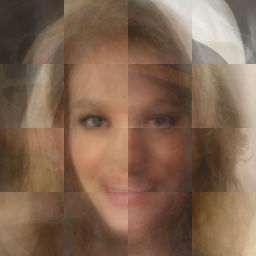}    \includegraphics[width=0.09\textwidth]{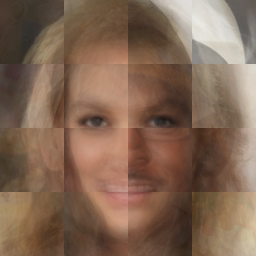}    \includegraphics[width=0.09\textwidth]{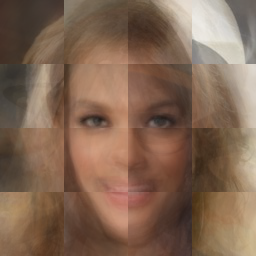}    \includegraphics[width=0.09\textwidth]{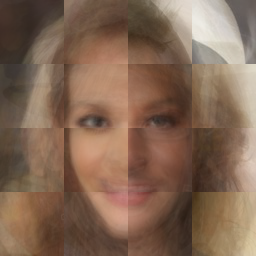}    \includegraphics[width=0.09\textwidth]{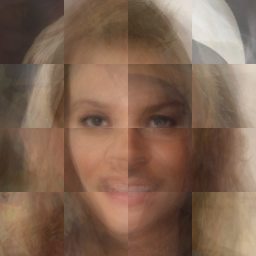}
    \caption{An in-place mosaic of the last image in the CelebA-HQ dataset using the first 1000 images, using only 10 warmup samples.}
    \label{fig:celeba_inplace_lowcompute}

\end{figure}

\begin{figure}
    \centering
    \includegraphics[width=0.09\textwidth]{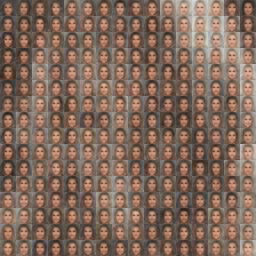}
    \includegraphics[width=0.09\textwidth]{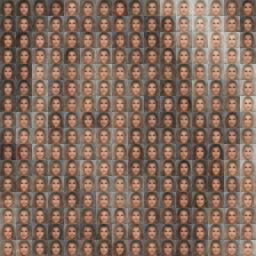}    \includegraphics[width=0.09\textwidth]{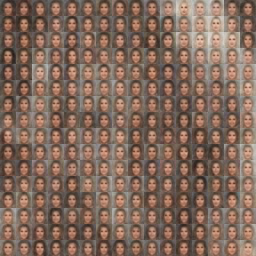}    \includegraphics[width=0.09\textwidth]{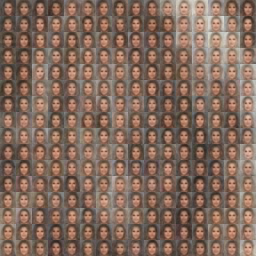}    \includegraphics[width=0.09\textwidth]{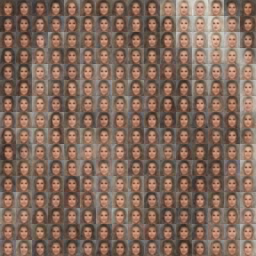}
    \includegraphics[width=0.09\textwidth]{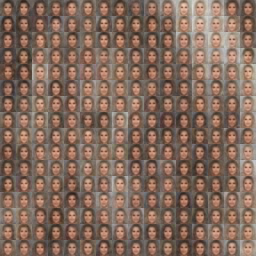}    \includegraphics[width=0.09\textwidth]{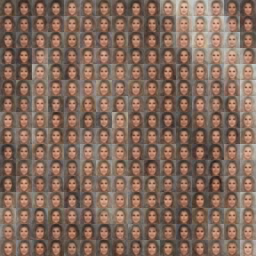}    \includegraphics[width=0.09\textwidth]{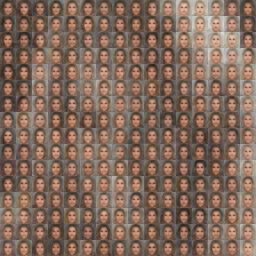}    \includegraphics[width=0.09\textwidth]{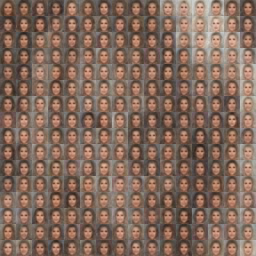}    \includegraphics[width=0.09\textwidth]{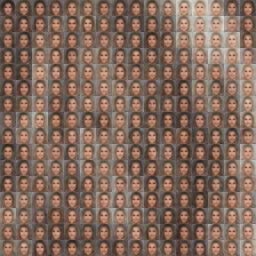}
    \caption{A photographic mosaic of the last image in the CelebA-HQ dataset using the first 1000 images downscaled to 16x16, using only 10 warmup samples.}
    \label{fig:celeba_photographic_lowcompute}
\end{figure}

We further note that while all previous results reported in this paper are with fully converged MCMC chains, we observed good qualitative results much earlier in the chain, sometimes after only a few steps. Artists using these tools who are only concerned with qualitative output may be able to achieve usable results in only a few minutes.

We show an extremely low compute in-place mosaic in Figure~\ref{fig:celeba_inplace_lowcompute} and an extremely low compute photographic mosaic in Figure~\ref{fig:celeba_photographic_lowcompute}. Both mosaics used only 10 warmup samples, whereas the results reported in Section~\ref{sec:inplace} used 1000 warmup samples; i.e. these mosaics were made using only 1\% of the compute relative to results reported earlier in the paper. Our extremely low compute in-place mosaic took 12 minutes and 36 seconds of wall clock time, and our extremely low compute photographic mosaic took 2 minutes and 41 seconds of wall clock time.

While in both of these experiments the MCMC chain has not fully converged, or even seen all 1000 source images, the results are qualitatively very similar to those of the fully converged chain, although the fine details do not match the target image as well as those produced by the fully converged chain. These experiments show the potential for our method to be used by artists who need to create mosaics for a large number of target data or who only have access to a consumer desktop CPU.

We noticed similar results in the audio mosaicing domain as well, with samples available for inspection at \url{https://soundcloud.com/datamosaicing/sets/low-compute}. In this experiment, we used 10 warmup samples to create a mosaic over the closest 200 source samples to each fragment. This experiment took 4 minutes and 22 seconds to run, with the majority of time spent in preprocessing. Despite taking 0.3\% of the wall clock time compared to the results reported in Section~\ref{sec:audiomosaicing}, these samples are largely indistinguishable to humans from those which originated from the fully converged chain.

We note that these extremely low compute experiments were only possible through the use of HMC, which can allow for large jumps in state space; attempting similar experiments with random walk Metropolis-Hastings would produce samples that are not significantly different from samples from the prior (i.e., averages of uniformly random source data fragments).

\section{Ethics Statement}

While this paper explores mosaicing via simulation-based inference as a tool for the algorithmic art community, techniques described in this paper could also be used to create convincing synthetic media, also known as ``deepfakes.'' As an example, a malicious actor could use a dataset of speech from a politician as source data, and a scandalous utterance (even recorded by the malicious actor themself) as the target datum. A ``mosaic'' of the target datum using the source data in this instance may appear to sound like a recording of the politician saying the target utterance. Furthermore, unlike deep learning-based methods, which typically require prohibitive amounts of data or compute to create, our algorithm works in low data and compute regimes, allowing for these methods to be used even by unsophisticated malicious actors.

While experiments on simulation-based inference to re-construct human speech for malicious purposes, or other malicious uses of the techniques described here, are out of the scope of this paper, we urge researchers to determine the fidelity with which similar simulation-based inference models could be used to create synthetic media, as well as further methods to detect such media. We note that likelihood-based generative models can also serve as generative classifiers~\cite{GambardellaThesis} and that it is possible to obtain likelihoods~\cite{Boelts2022} or likelihood-to-evidence ratios from stochastic simulators~\cite{Hermans2020}, and thus any such model that creates synthetic media may also be able to be used to detect synthetic media created by that model or similar models.

\section{Conclusion and Future Work}

In this work, we have introduced a simple model for mosaicing through the simulation-based inference paradigm and explored applications of this model in algorithmic art. The simplicity of the model made it amenable to parallelization as well as allowed for its use in a data-agnostic manner. Despite this simplicity, we demonstrated successful experiments over multiple modalities, even in extremely low compute scenarios.

Our model is amenable to further modification and experimentation, such as by giving the model data in different modalities, giving the model data with different fragmentations, or even through modification of the model itself by adding more latent variables or changing the exact details of how data fragments can be combined. We believe that with recent developments such as the wide availability of HMC in standard probabilistic programming libraries, the simulation-based inference paradigm will grow to be a valuable tool for the algorithmic art community, and we offer this work as evidence for its usefulness in data mosaicing.

\clearpage
\bibliographystyle{named}
\bibliography{references}

\begin{thebibliography}{}

\bibitem[\protect\citeauthoryear{Baydin \bgroup \em et al.\egroup
  }{2019a}]{Baydin2018}
Atılım~Güneş Baydin, Lukas Heinrich, Wahid Bhimji, Bradley Gram-Hansen,
  Gilles Louppe, Lei Shao, P.~Prabhat, Kyle Cranmer, and Frank Wood.
\newblock {Efficient probabilistic inference in the quest for physics beyond
  the standard model}.
\newblock In {\em Advances in Neural Information Processing Systems},
  volume~33, 2019.

\bibitem[\protect\citeauthoryear{Baydin \bgroup \em et al.\egroup
  }{2019b}]{Baydin2019}
Atılım~Güneş Baydin, Lei Shao, Wahid Bhimji, Lukas Heinrich, Lawrence
  Meadows, Jialin Liu, Andreas Munk, Saeid Naderiparizi, Bradley Gram-Hansen,
  Gilles Louppe, Mingfei Ma, Xiaohui Zhao, Philip Torr, Victor Lee, Kyle
  Cranmer, {Prabhat}, and Frank Wood.
\newblock {Etalumis: Bringing probabilistic programming to scientific
  simulators at scale}.
\newblock In {\em International Conference for High Performance Computing,
  Networking, Storage and Analysis, SC}, 2019.

\bibitem[\protect\citeauthoryear{Bingham \bgroup \em et al.\egroup
  }{2019}]{bingham2019pyro}
Eli Bingham, Jonathan~P. Chen, Martin Jankowiak, Fritz Obermeyer, Neeraj
  Pradhan, Theofanis Karaletsos, Rohit Singh, Paul Szerlip, Paul Horsfall, and
  Noah~D. Goodman.
\newblock {Pyro: Deep Universal Probabilistic Programming}.
\newblock {\em Journal of Machine Learning Research}, 20(1):973--978, 2019.

\bibitem[\protect\citeauthoryear{Boelts \bgroup \em et al.\egroup
  }{2022}]{Boelts2022}
Jan Boelts, Jan-Matthis Lueckmann, Richard Gao, and Jakob~H Macke.
\newblock {Flexible and efficient simulation-based inference for models of
  decision-making}.
\newblock {\em eLife}, 11, 7 2022.

\bibitem[\protect\citeauthoryear{Bradbury \bgroup \em et al.\egroup
  }{2018}]{jax2018github}
James Bradbury, Roy Frostig, Peter Hawkins, Matthew~James Johnson, Chris Leary,
  Dougal Maclaurin, George Necula, Adam Paszke, Jake VanderPlas, Skye
  Wanderman-Milne, and Qiao Zhang.
\newblock {JAX: composable transformations of Python+NumPy programs}, 2018.

\bibitem[\protect\citeauthoryear{Brooks and Gelman}{1998}]{Brooks1998}
Stephen~P. Brooks and Andrew Gelman.
\newblock {General Methods for Monitoring Convergence of Iterative
  Simulations}.
\newblock {\em Journal of Computational and Graphical Statistics}, 7(4):434,
  1998.

\bibitem[\protect\citeauthoryear{Carroll}{1871}]{Carroll1871}
Lewis Carroll.
\newblock {\em {Through the Looking-Glass, and What Alice Found There}}.
\newblock Macmillan, 1871.

\bibitem[\protect\citeauthoryear{Coleman \bgroup \em et al.\egroup
  }{2010}]{ColemanMosaic}
Graham Coleman, Esteban Maestre, and Jordi Bonada.
\newblock {Augmenting sound mosaicing with descriptor-driven transformation}.
\newblock In {\em Proceedings of DAFx}, 2010.

\bibitem[\protect\citeauthoryear{Cranmer \bgroup \em et al.\egroup
  }{2020}]{cranmer2020frontier}
Kyle Cranmer, Johann Brehmer, and Gilles Louppe.
\newblock {The frontier of simulation-based inference}.
\newblock {\em Proceedings of the National Academy of Sciences},
  117(48):30055--30062, 2020.

\bibitem[\protect\citeauthoryear{Davidson}{1986}]{DavidsonEpitaphs}
Donald Davidson.
\newblock {A Nice Derangement of Epitaphs}.
\newblock In {\em Truth and Interpretation}, pages 433--446. Blackwell, 1986.

\bibitem[\protect\citeauthoryear{Deleuze and Guattari}{1980}]{Mille1980}
Gilles Deleuze and Félix Guattari.
\newblock {\em {Mille plateaux}}.
\newblock Les {\'{E}}ditions de Minuit, 1980.

\bibitem[\protect\citeauthoryear{Driedger \bgroup \em et al.\egroup
  }{2015}]{LetItBee2015}
Jonathan Driedger, Thomas Pr{\"{a}}tzlich, and Meinard M{\"{u}}ller.
\newblock {Let it bee – Towards NMF-inspired audio mosaicing}.
\newblock In {\em Proceedings of the 16th International Society for Music
  Information Retrieval Conference, ISMIR 2015}, 2015.

\bibitem[\protect\citeauthoryear{Duane \bgroup \em et al.\egroup
  }{1987}]{DuaneHMC}
Simon Duane, A.~D. Kennedy, Brian~J. Pendleton, and Duncan Roweth.
\newblock {Hybrid Monte Carlo}.
\newblock {\em Physics Letters B}, 195(2), 1987.

\bibitem[\protect\citeauthoryear{Finkelstein and
  Range}{1998}]{finkelstein1998image}
Adam Finkelstein and Marisa Range.
\newblock {Image Mosaics}.
\newblock In {\em International conference on raster imaging and digital
  typography}, pages 11--22. Springer, 1998.

\bibitem[\protect\citeauthoryear{Gambardella}{2021}]{GambardellaThesis}
Andrew Gambardella.
\newblock {\em {Deep transfer learning with Bayesian inference}}.
\newblock PhD thesis, University of Oxford, 2021.

\bibitem[\protect\citeauthoryear{Gelman and Rubin}{1992}]{Gelman1992}
Andrew Gelman and Donald~B. Rubin.
\newblock {Inference from iterative simulation using multiple sequences}.
\newblock {\em Statistical Science}, 7(4):457--472, 1992.

\bibitem[\protect\citeauthoryear{Hastings}{1970}]{Hastings1970}
W.~K. Hastings.
\newblock {Monte Carlo Sampling Methods Using Markov Chains and Their
  Applications}.
\newblock {\em Biometrika}, 57(1):97, 1970.

\bibitem[\protect\citeauthoryear{Hermans \bgroup \em et al.\egroup
  }{2020}]{Hermans2020}
Joeri Hermans, Volodimir Begy, and Gilles Louppe.
\newblock {Likelihood-free MCMC with amortized approximate ratio estimators}.
\newblock In {\em 37th International Conference on Machine Learning, ICML
  2020}, volume PartF168147-6, 2020.

\bibitem[\protect\citeauthoryear{Hoffman and Gelman}{2014}]{HoffmanNUTS}
Matthew~D. Hoffman and Andrew Gelman.
\newblock {The no-U-turn sampler: Adaptively setting path lengths in
  Hamiltonian Monte Carlo}.
\newblock {\em Journal of Machine Learning Research}, 15, 2014.

\bibitem[\protect\citeauthoryear{Hoffman \bgroup \em et al.\egroup
  }{2009}]{HoffmanSpectral}
Matthew~D. Hoffman, Perry~R. Cook, and David~M. Blei.
\newblock {Bayesian spectral matching: Turning Young MC into MC Hammer via MCMC
  sampling}.
\newblock In {\em Proceedings of the 2009 International Computer Music
  Conference, ICMC 2009}, 2009.

\bibitem[\protect\citeauthoryear{Karras \bgroup \em et al.\egroup
  }{2018}]{ProgressiveGANs}
Tero Karras, Timo Aila, Samuli Laine, and Jaakko Lehtinen.
\newblock {Progressive growing of GANs for improved quality, stability, and
  variation}.
\newblock In {\em 6th International Conference on Learning Representations,
  ICLR 2018 - Conference Track Proceedings}, 2018.

\bibitem[\protect\citeauthoryear{Klein \bgroup \em et al.\egroup
  }{2002}]{KleinVideoMosaics}
Allison~W. Klein, Tyler Grant, Adam Finkelstein, and Michael~F. Cohen.
\newblock {Video mosaics}.
\newblock In {\em Proceedings of the second international symposium on
  Non-photorealistic animation and rendering - NPAR '02}, page~21, New York,
  New York, USA, 2002. ACM Press.

\bibitem[\protect\citeauthoryear{Lazier and Cook}{2003}]{Mosievius}
Ari Lazier and Perry Cook.
\newblock {Mosievius: Feature Driven Interactive Audio Mosaicing}.
\newblock {\em Audio}, 2003.

\bibitem[\protect\citeauthoryear{Marjoram \bgroup \em et al.\egroup
  }{2003}]{marjoram2003markov}
Paul Marjoram, John Molitor, Vincent Plagnol, and Simon Tavar{\'{e}}.
\newblock {Markov chain Monte Carlo without likelihoods}.
\newblock {\em Proceedings of the National Academy of Sciences of the United
  States of America}, 100(26):15324--15328, 2003.

\bibitem[\protect\citeauthoryear{Metropolis \bgroup \em et al.\egroup
  }{1953}]{Metropolis1953}
Nicholas Metropolis, Arianna~W. Rosenbluth, Marshall~N. Rosenbluth, Augusta~H.
  Teller, and Edward Teller.
\newblock {Equation of state calculations by fast computing machines}.
\newblock {\em Journal of Chemical Physics}, 21(6):1087--1092, 1953.

\bibitem[\protect\citeauthoryear{Neal}{1996}]{RadfordHMC}
Radford~M. Neal.
\newblock {Monte Carlo Implementation}.
\newblock In {\em Bayesian Learning for Neural Networks}, pages 55--98.
  Springer, 1996.

\bibitem[\protect\citeauthoryear{Phan \bgroup \em et al.\egroup
  }{2019}]{phan2019composable}
Du~Phan, Neeraj Pradhan, and Martin Jankowiak.
\newblock {Composable Effects for Flexible and Accelerated Probabilistic
  Programming in NumPyro}.
\newblock In {\em NeurIPS 2019 Program Transformations for Machine Learning
  Workshop}, 12 2019.

\bibitem[\protect\citeauthoryear{Sejdi{\'{c}} \bgroup \em et al.\egroup
  }{2009}]{SignalProcessingAdvances}
Ervin Sejdi{\'{c}}, Igor Djurovi{\'{c}}, and Jin Jiang.
\newblock {Time-frequency feature representation using energy concentration: An
  overview of recent advances}.
\newblock {\em Digital Signal Processing: A Review Journal}, 19(1), 2009.

\bibitem[\protect\citeauthoryear{van~de Meent \bgroup \em et al.\egroup
  }{2018}]{van2018introduction}
Jan-Willem van~de Meent, Brooks Paige, Hongseok Yang, and Frank Wood.
\newblock {An Introduction to Probabilistic Programming}.
\newblock {\em arXiv preprint}, 2018.

\bibitem[\protect\citeauthoryear{Wilkinson}{2013}]{wilkinson2013approximate}
Richard~David Wilkinson.
\newblock {Approximate Bayesian computation (ABC) gives exact results under the
  assumption of model error}.
\newblock {\em Statistical Applications in Genetics and Molecular Biology},
  12(2):129--141, 2013.

\bibitem[\protect\citeauthoryear{Wingate \bgroup \em et al.\egroup
  }{2011}]{wingate2011lightweight}
David Wingate, Andreas Stuhlm{\"{u}}ller, and Noah~D. Goodman.
\newblock {Lightweight implementations of probabilistic programming languages
  via transformational compilation}.
\newblock In {\em Journal of Machine Learning Research}, volume~15, pages
  770--778, 2011.

\bibitem[\protect\citeauthoryear{Wittgenstein}{1953}]{WittgensteinInvestigations}
Ludwig Wittgenstein.
\newblock {\em {Philosophische Untersuchungen}}.
\newblock Blackwell, 1953.

\bibitem[\protect\citeauthoryear{Zils and Pachet}{2001}]{MusicalMosaicing}
Aymeric Zils and François Pachet.
\newblock {Musical mosaicing}.
\newblock {\em Digital Audio Effects (DAFx)}, 2001.

\end{thebibliography}

\end{document}